\documentclass[conference]{IEEEtran}
\IEEEoverridecommandlockouts
\usepackage{cite}
\usepackage{amsmath,amssymb,amsfonts}
\usepackage{algorithmic}
\usepackage{graphicx}
\usepackage{textcomp}
\usepackage{xcolor}
\usepackage{url}
\def\BibTeX{{\rm B\kern-.05em{\sc i\kern-.025em b}\kern-.08em
    T\kern-.1667em\lower.7ex\hbox{E}\kern-.125emX}}
\begin{document}

\title{Colepp: uma ferramenta multiplataforma para coleta de dados de dispositivos vestíveis}

\author{\IEEEauthorblockN{Vinicius Moraes de Jesus}
\IEEEauthorblockA{\textit{Universidade Federal do Espírito Santo} \\
Vitória-ES, Brasil \\
vinicius.jesus@edu.ufes.br}
\and
\IEEEauthorblockN{Andre Georghton Cardoso Pacheco}
\IEEEauthorblockA{\textit{Universidade Federal do Espírito Santo} \\
Vitória-ES, Brasil \\
apacheco@inf.ufes.br}
}

\maketitle

\begin{abstract}
  The widespread adoption of wearable devices such as smartwatches and fitness trackers has fueled the demand for reliable physiological and movement data collection tools. However, challenges such as limited access to large, high-quality public datasets and a lack of control over data collection conditions hinder the development of robust algorithms. This work presents Colepp, an open-source, cross-platform tool designed to collect and synchronize data from multiple wearable devices, including heart rate (via ECG and PPG) and motion signals (accelerometer and gyroscope). The system integrates a smartphone as a central hub, receiving data from a Polar H10 chest strap and a Wear OS smartwatch, and exporting synchronized datasets in CSV format. Through a custom synchronization protocol and user-friendly interface, Colepp facilitates the generation of customizable, real-world datasets suitable for applications such as human activity recognition and heart rate estimation. A use case shows the effectiveness of the tool in producing consistent and synchronized signals.
\end{abstract}

\begin{IEEEkeywords}
Wearable,  Data collection,  Heart rate monitoring,  Photoplethysmography (PPG),  Electrocardiogram (ECG),  Human activity recognition (HAR),  Data synchronization,  Open-source tool.
\end{IEEEkeywords}

\section{Introdução}

O uso de dispositivos vestíveis (\textit{wearables}) tem crescido significativamente ao longo dos últimos anos. Com os avanços tecnológicos e a crescente preocupação com saúde e bem-estar, esses dispositivos, como \textit{smartwatches} e pulseiras fitness, tornaram-se cada vez mais populares entre os consumidores \cite{AndreP2024}. Prevê-se que o número global de usuários no segmento \textit{smartwatches} do mercado de saúde digital aumente continuamente entre 2024 e 2029, totalizando 285,8 milhões de usuários, um crescimento de 62,86\% \cite{wearable-statistc}. 

Dentre as principais aplicações destes dispositivos, destaca-se a medição da frequência cardíaca (FC) e a análise de movimentos, possibilitadas por sensores avançados presentes nestes dispositivos. A FC é um sinal vital de saúde que pode ser obtido por estes dispositivos através dos sensores de eletrocardiograma (ECG) e fotopletismografia (PPG). A combinação de não ser invasivo, facilidade de uso e variedade de preços do PPG fez a indústria de relógios inteligentes vestíveis florescer em alguns anos \cite{ThakurSmritis2023}, o tornando o principal sensor para medição de FC em dispositivos desse tipo. A coleta desses dados é essencial para impulsionar o desenvolvimento técnico e científico, fornecendo insumos para a criação e aprimoramento de novos algoritmos como os propostos por Pacheco et al. \cite{AndreP2024} e Hassan et. al \cite{Hasan2023}.

A PPG é uma técnica óptica não invasiva que mede mudanças no volume sanguíneo, sendo comumente empregada em \textit{smartwatches} por sua facilidade de monitoramento contínuo do sistema cardiovascular durante as atividades diárias \cite{allen2007photoplethysmography}. No entanto, a precisão do sinal PPG é significativamente afetada por artefatos de movimento (do inglês: \textit{motion artifacts}) \cite{motion-artifacts}.
Em virtude disso, um problema recorrente na construção de datasets relacionados a \textit{wearables} é a falta de controle e transparência sobre as condições em que os dados foram coletados, devido à dificuldade de rotular um conjunto de dados que possa capturar as frequências cardíacas em diferentes atividades humanas \cite{IsmailShahid2021}. Essa limitação compromete a generalização e a robustez dos modelos treinados, especialmente quando se busca desenvolver algoritmos aplicáveis em contextos do mundo real.

Além disso, é comum que grandes bases de dados na área sejam privadas, como a descrita em Pacheco et al. \cite{Pacheco2023}, que reúne mais de 180 horas de dados com mais de 800 sessões no total. Em contrapartida, bases públicas disponíveis costumam ter amostras reduzidas e protocolos menos padronizados. A base de Jarchi et al. \cite{Jarchi2017}, por exemplo, inclui apenas 8 voluntários com 24 sessões no total. Já a base de Biagetti et al. \cite{Biagetti2020} conta com 7 participantes e 35 sessões.

Esse tipo de limitação evidencia a necessidade de ferramentas que permitam a coleta de dados em condições controláveis e com protocolos customizáveis. A geração de dados de qualidade é um passo fundamental para o desenvolvimento de algoritmos confiáveis em dispositivos \textit{wearables}, especialmente em aplicações sensíveis como monitoramento de saúde ou detecção de atividades. 

Neste contexto, esse trabalho apresenta uma ferramenta multiplataforma e \textit{open source}\footnote{\url{https://github.com/life-ufes/Colepp}}
, sob os termos da licença MIT, capaz de coletar dados de sinais vitais e movimentos de usuários a partir de diferentes dispositivos, fazendo a sincronização desses dados e organizando-os em um formato adequado para a criação de conjuntos de dados, foi disponibilizado um vídeo tutorial da ferramenta\footnote{\url{https://drive.google.com/file/d/1TlWn2WMI4MvkFQ_k9JsjuzJQwiJQs4Qa/view}}. Desse modo, a criação de ferramentas como o \textit{framework} proposto neste trabalho é importante, pois permite a coleta estruturada, sincronizada e personalizável de múltiplos sinais fisiológicos e de movimento. O sistema visa facilitar a geração de dados reais e personalizados, viabilizando pesquisas e aplicações em áreas como detecção de atividades humanas (HAR) \cite{dentamaro2024human} e estimativa de frequência cardíaca personalizada \cite{AndreP2024}. Com isso, pesquisadores e desenvolvedores podem gerar seus próprios conjuntos de dados, ajustando parâmetros experimentais, cenários e perfis de usuários, contribuindo para modelos mais precisos e generalizáveis.

\begin{figure*}[h]
    \centering
    \includegraphics[width=0.65\linewidth]{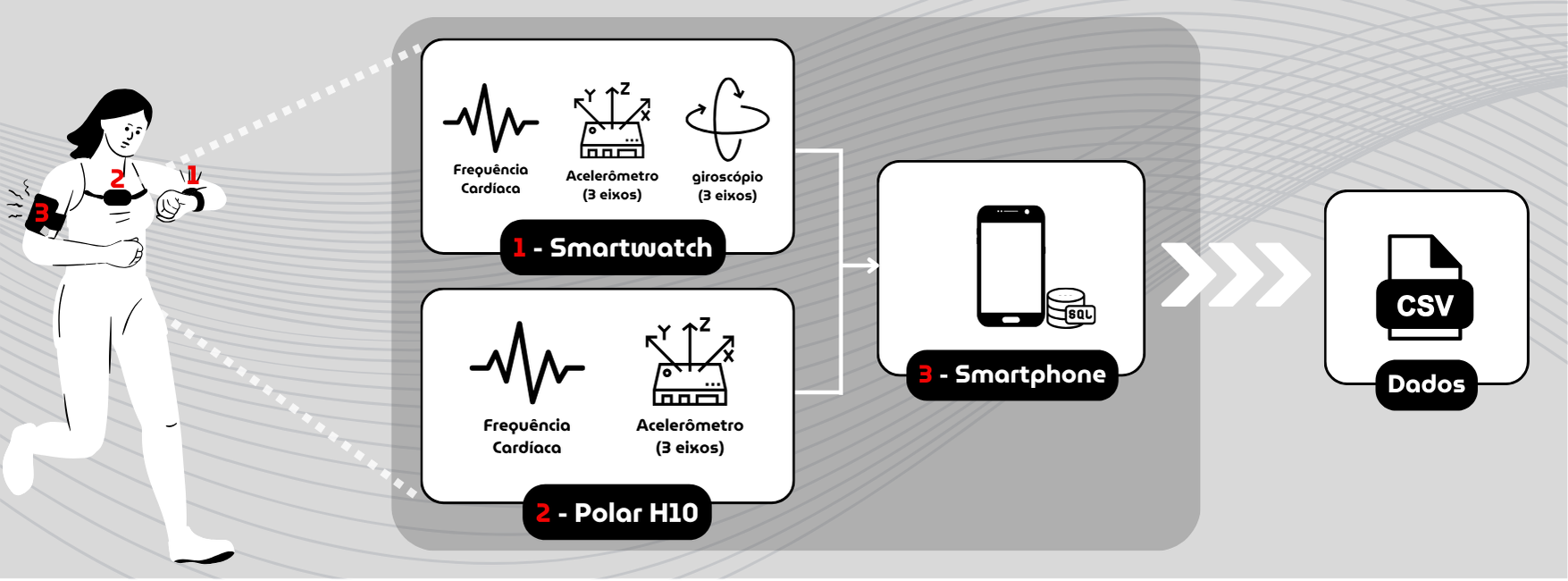}
    \caption{Uma visão geral do funcionamento da ferramenta. Observe que se trata de um sistema multiplataforma que coleta e sincroniza dados fisiológicos e de movimento em tempo real, centralizando-os no \textit{smartphone} para exportação em CSV.}
    \label{fig:overview}
\end{figure*}

\section{Ferramenta}

Nesta seção é detalhado o desenvolvimento da ferramenta, incluindo a visão geral do sistema, suas funcionalidades, o processo de sincronização de dados e as tecnologias utilizadas.

\subsection{Visão geral}

A ferramenta proposta neste trabalho consiste em um sistema multiplataforma voltado para a coleta sincronizada de dados fisiológicos e de movimento, conforme representado na Figura \ref{fig:overview}. Para a coleta, é necessário que o voluntário utilize simultaneamente três dispositivos: uma cinta Polar H10\footnote{Cinta peitoral para monitoramento cardíaco equipada com sensor de ECG.}, um \textit{smartwatch} com \textit{Wear OS}\footnote{Relógio inteligente com sistema operacional Wear OS e múltiplos sensores integrados.} e um \textit{smartphone} Android\footnote{Telefone celular com sistema Android.}. O \textit{smartphone} atua como unidade central de controle e armazenamento, sendo responsável por receber os dados dos sensores de FC e de movimentos do Polar H10 e do \textit{smartwatch}. Importante destacar que o \textit{smartphone} não precisa estar fisicamente junto ao corpo do voluntário durante a coleta, desde que permaneça no alcance de comunicação dos dispositivos\footnote{Por exemplo, se a coleta for realizada em um ambiente indoor em uma esteira, o smartphone pode ser controlado inteiramente por um agente externo (ex: educador físico)}. Todos os dados recebidos são armazenados localmente em um banco de dados relacional e podem ser exportados sob demanda no formato CSV (\textit{Comma-Separated Values}), facilitando seu uso em análises posteriores e compartilhamento com outros pesquisadores.

\begin{sloppypar}
A ferramenta é composta por dois aplicativos. O primeiro foi desenvolvido nativamente para Android\footnote{Sistema operacional \textit{open source} mantido pela Google: \url{https://www.android.com}} e deve ser instalado no \textit{smartphone}. Esse aplicativo é responsável por estabelecer a conexão com a cinta Polar H10 e com o \textit{smartwatch}, recebendo os dados provenientes dos dispositivos, realizando sua sincronização, armazenando-os localmente e, quando solicitado, exportando-os em formato CSV para fins de compartilhamento. O segundo aplicativo foi desenvolvido para \textit{Wear OS} e é executado diretamente no \textit{smartwatch}, sendo encarregado de coletar os dados dos sensores e transferi-los para o \textit{smartphone}. Vale destacar que, para a cinta Polar H10, não foi necessário desenvolver um aplicativo dedicado, pois a fabricante disponibiliza um SDK\footnote{\url{https://github.com/polarofficial/polar-ble-sdk}} (Kit de Desenvolvimento de Software) que inclui uma API (Interface de Programação de Aplicações) para acesso aos dados do dispositivo via BLE (\textit{Bluetooth Low Energy}).
\end{sloppypar}

\begin{figure}
    \centering
    \includegraphics[width=0.72\linewidth]{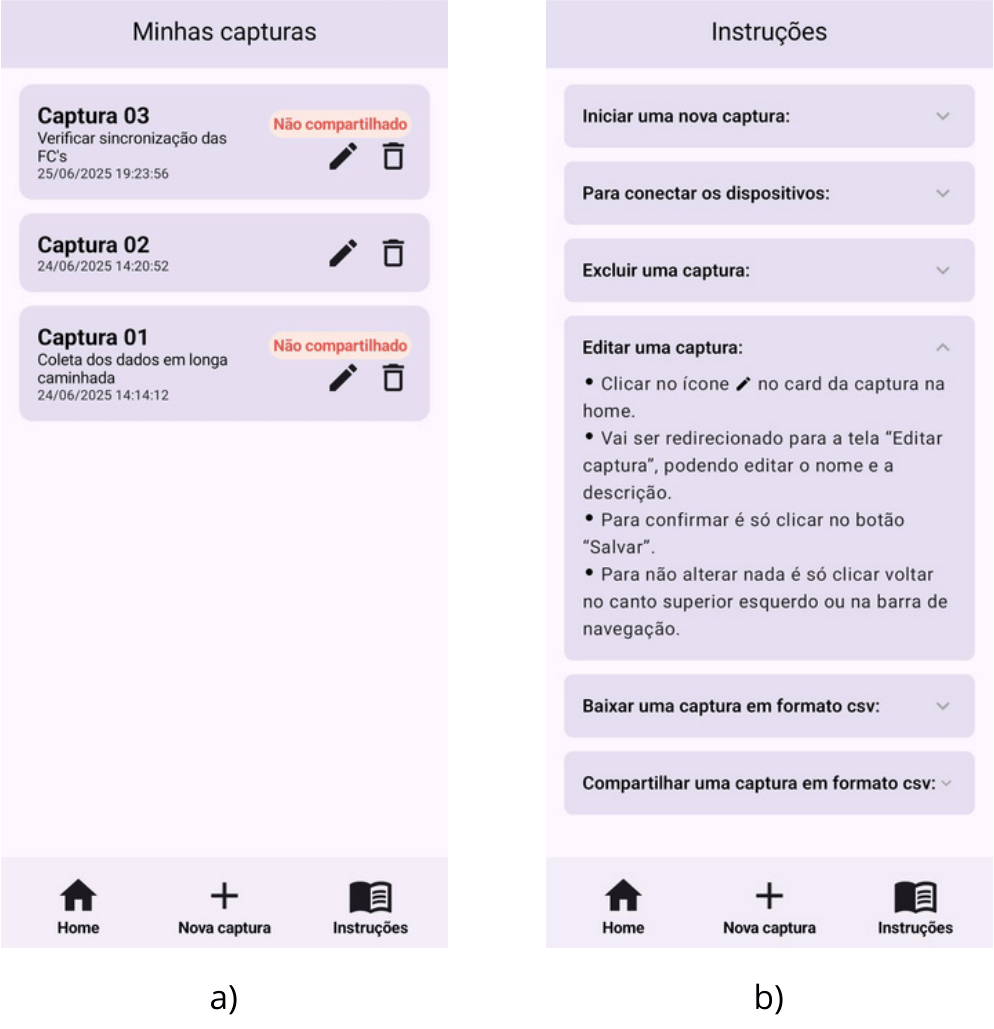}
    \caption{Interface inicial do aplicativo no \textit{smartphone}. À esquerda, a tela principal exibe as opções para iniciar uma nova coleta ou visualizar o histórico. À direita, a tela de instruções orienta o usuário sobre a utilização do aplicativo.}
    \label{fig:home_and_instructions}
\end{figure}

\begin{figure}
    \centering
    \includegraphics[width=0.72\linewidth]{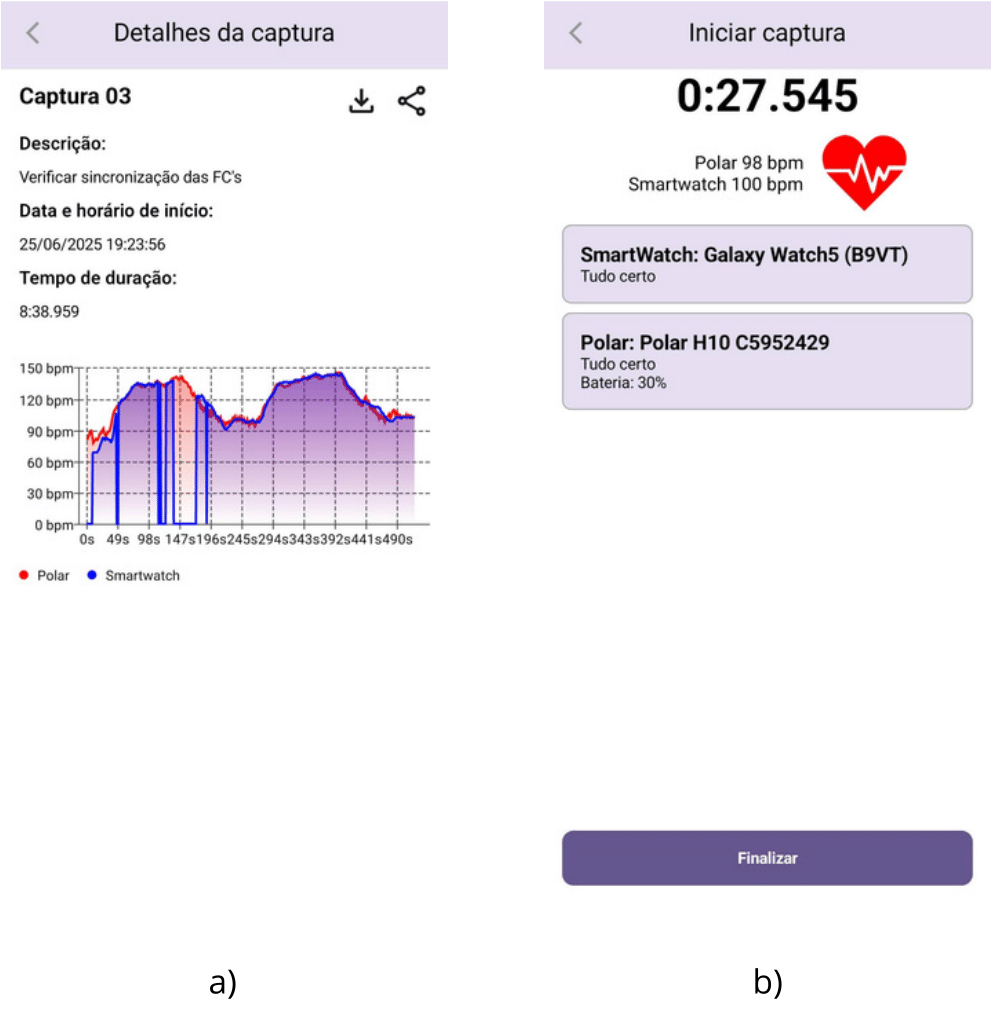}
    \caption{Visualização detalhada de uma sessão de coleta no aplicativo do \textit{smartphone}. À esquerda, a tela exibe os metadados da sessão e a evolução da frequência cardíaca (FC). À direita, a tela de coleta em tempo real apresenta status de conexão, valores de FC dos dispositivos e tempo decorrido.}
    \label{fig:details_and_recording}
\end{figure}

\begin{figure}
    \centering
    \includegraphics[width=0.65\linewidth]{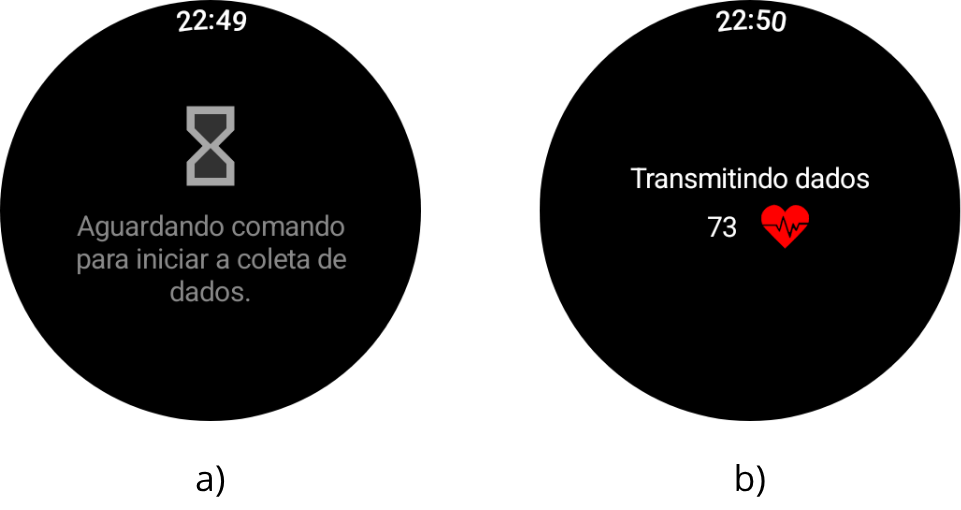}
    \caption{Interface do aplicativo no \textit{smartwatch}. À esquerda, a tela indica que o dispositivo está aguardando o início da coleta. À direita, a tela ativa mostra o valor da frequência cardíaca estimada por PPG e o status de coleta em andamento.}
    \label{fig:smartwatch_screens}
\end{figure}

\subsection{Funcionalidades}

A ferramenta desenvolvida possui três funcionalidades principais: (i) coleta simultânea de dados de dois dispositivos vestíveis (\textit{wearables}), (ii) sincronização temporal precisa entre os dados coletados e (iii) armazenamento e exportação dos dados no formato CSV para posterior análise ou compartilhamento.

Durante uma sessão de coleta, o Polar H10 registra valores do acelerômetro e da frequência cardíaca (FC), estimada por meio da técnica de eletrocardiograma (ECG). Já o \textit{smartwatch} fornece dados do acelerômetro, do giroscópio e também da FC, cuja estimativa é realizada utilizando a técnica de fotopletismografia (PPG).

A presença de dois sensores de FC baseados em técnicas distintas possibilita comparações de desempenho e estudos de confiabilidade entre as tecnologias, sendo que a FC extraída via ECG é normalmente utilizada como referência para o desenvolvimento de técnicas baseadas em PPG \cite{AndreP2024}.

A interface do aplicativo no \textit{smartphone} foi projetada para facilitar a usabilidade e o acompanhamento da coleta em tempo real. Ela inclui uma tela inicial (Figura \ref{fig:home_and_instructions}a), que permite iniciar uma nova coleta ou acessar o histórico; uma tela de instruções (Figura \ref{fig:home_and_instructions}b), que orienta o usuário quanto ao uso correto dos dispositivos; uma tela de detalhes da coleta (Figura \ref{fig:details_and_recording}a), que apresenta título, descrição, data, duração da sessão e um gráfico com a variação da FC de ambos os dispositivos ao longo do tempo; e uma tela de coleta em andamento (Figura \ref{fig:details_and_recording}b), que exibe o botão de início da captura, o status de conexão de cada dispositivo, as FC registradas em tempo real e o tempo total decorrido.

O aplicativo desenvolvido para o \textit{smartwatch} possui uma interface simples e funcional, composta por duas telas. A primeira, de espera (Figura \ref{fig:smartwatch_screens}a), informa que o dispositivo está pronto e conectado, aguardando o comando do \textit{smartphone} para iniciar a captura. A segunda, de coleta em andamento (Figura \ref{fig:smartwatch_screens}b), exibe em tempo real a FC estimada e um indicativo visual de captura ativa, garantindo simplicidade e foco na correta aquisição dos dados.

É importante destacar que, durante os experimentos, os dados de FC do \textit{smartwatch} ocasionalmente registram valores iguais a zero. Isso ocorre, pois quando a qualidade do sinal do PPG é muito ruim, para evitar uma estimativa muito errada, o dispositivo apresenta a FC do usuário igual a zero. Logo, este comportamento não indica uma falha de conexão entre os dispositivos.

\subsubsection{Sincronização} 

A sincronização é uma característica importante desta ferramenta, permitindo alinhar temporalmente os dados coletados pelo \textit{smartwatch} em relação aos obtidos pelo Polar H10. Os dados provenientes do \textit{smartwatch} são acompanhados por marcações de tempo em nanosegundos, baseadas no tempo decorrido desde a inicialização do dispositivo. Por outro lado, os dados do acelerômetro do Polar utilizam uma marcação de tempo também em nanosegundos, porém com referência ao tempo transcorrido desde 1º de janeiro de 2000, 00:00:00 UTC. Já os dados de batimento cardíaco fornecidos pelo Polar não possuem marcação temporal associada ao dispositivo; nesse caso, é registrado o instante em que os dados são recebidos pelo \textit{smartphone}; além do tempo de início e término da coleta, tanto em nanosegundos desde a inicialização do sistema quanto em milissegundos no formato \textit{Unix time} (tempo decorrido desde 1º de janeiro de 1970, 00:00:00 UTC). Embora todas essas informações sejam fundamentais para a sincronização, elas ainda não são suficientes para estabelecer uma referência de tempo comum entre o \textit{smartphone} e o \textit{smartwatch}. Para suprir essa lacuna, foi implementado um protocolo de troca de mensagens assíncronas utilizando a API MessageClient\footnote{\url{https://developer.android.com/training/wearables/data/overview}}.

\begin{figure*}[h]
    \centering
    \includegraphics[width=0.73\linewidth]{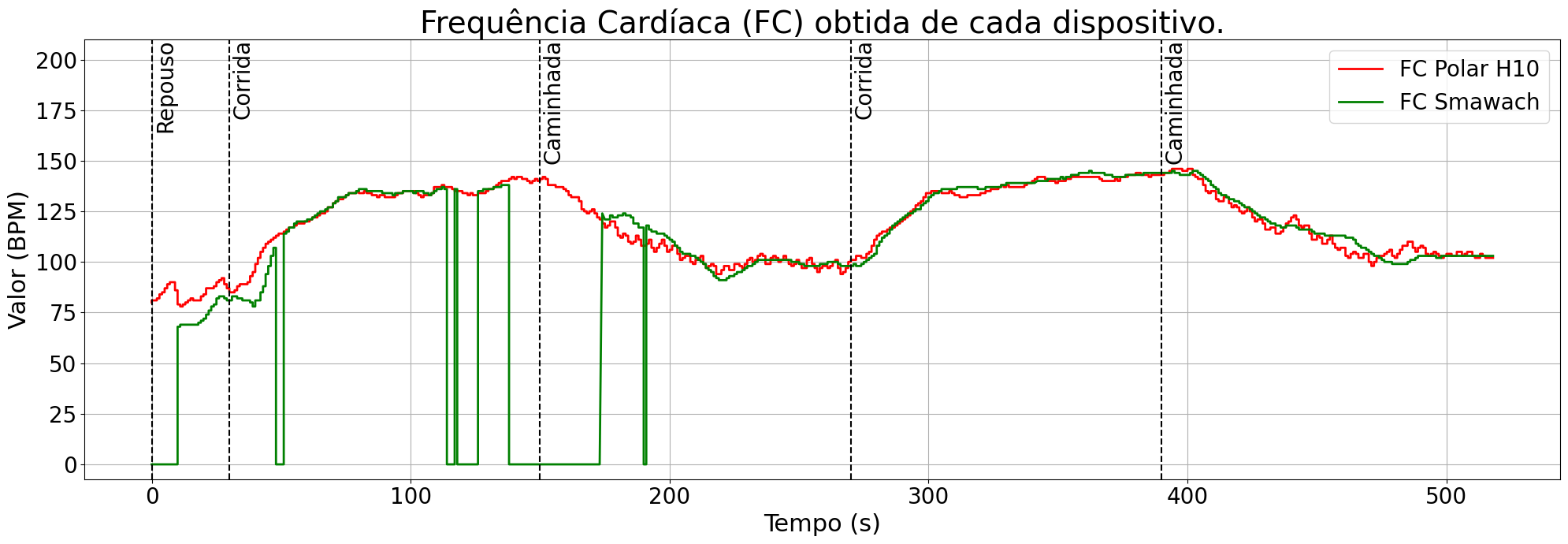}
    \caption{Frequência Cardíaca (FC) obtida simultaneamente pelo Polar H10 e pelo \textit{smartwatch} com Wear OS durante as diferentes fases do protocolo de coleta.}
    % \Description{}
    \label{fig:hr_devices}
\end{figure*}

O processo consiste em troca de mensagens entre o \textit{smartphone} e o \textit{smartwatch}, contendo um carimbo de tempo em nanosegundos decorridos desde a inicialização do dispositivo. Sendo a sequência: (i) o \textit{smartphone} envia uma mensagem contendo o carimbo de tempo $t_1$; (ii) o \textit{smartwatch} registra seu próprio carimbo de tempo $t_2$ e envia uma mensagem contendo os valores de $t_1$ e $t_2$; (iii) o \textit{smartphone}, ao receber a resposta, registra o carimbo de tempo $t_3$. Com os três valores ($t_1$, $t_2$ e $t_3$), é possível estimar a diferença de tempo (\textit{offset}) entre os dois dispositivos, utilizando a equação a seguir: 

\begin{equation}
offset = t_1 - t_2 - \left( \frac{t_3 - t_1}{2} \right)    
\end{equation}

Essa operação é repetida diversas vezes e, ao final do processo, calcula-se a média dos \textit{offsets}, resultando em uma estimativa mais precisa da diferença temporal entre os dispositivos. Com essa estimativa, possibilita-se converter os registros temporais do \textit{smartwatch} para a base de tempo do \textit{smartphone}. Para isso, basta somar o valor médio dos \textit{offsets} aos timestamps dos dados provenientes do \textit{smartwatch}, garantindo assim que todas as informações estejam alinhadas temporalmente em uma mesma referência de tempo.

\subsubsection{Tecnologias utilizadas}

A ferramenta foi desenvolvida para a plataforma Android utilizando a linguagem Kotlin\footnote{\url{https://kotlinlang.org}}, organizada em uma arquitetura modular composta por dois módulos principais: o módulo \textit{mobile} (aplicativo para \textit{smartphone}) e o módulo \textit{wear} (aplicativo para \textit{smartwatch}). O desenvolvimento seguiu o padrão arquitetural MVVM (\textit{Model-View-ViewModel}), promovendo uma separação clara de responsabilidades e facilitando a manutenção e escalabilidade do código. As interfaces de usuário foram construídas com Jetpack Compose\footnote{Biblioteca declarativa moderna do Android: \url{https://developer.android.com/compose}}. Para o armazenamento dos dados, foi utilizado o banco de dados local SQLite\footnote{\url{https://www.sqlite.org}}.

\subsection{Limitações}

A ferramenta apresenta algumas limitações decorrentes de dependências tecnológicas. A primeira delas é que o aplicativo do \textit{smartwatch} requer que o dispositivo utilize o sistema operacional \textit{Wear OS} na versão 3 ou superior, o que restringe a compatibilidade com dispositivos mais antigos. Além disso, não foi possível coletar os dados brutos do sensor de PPG, por conta de restrições impostas pela fabricante do dispositivo utilizado, a Samsung. O acesso direto às informações brutas do sensor requer parceria comercial homologada com a empresa, o que inviabilizou a extração dos dados brutos neste projeto. Porém, o código é adaptado para que, em caso de parceria, os dados possam ser facilmente coletados no dispositivo.

\section{Exemplo de caso de uso}

Para avaliar a qualidade dos dados e a eficácia do processo de sincronização, foi realizada uma coleta experimental para exemplificar um caso de uso da ferramenta. Um voluntário seguiu o seguinte protocolo de exercício: 30 segundos iniciais em repouso, seguidos por quatro fases de dois minutos cada, alternando entre corrida e caminhada, totalizando 8 minutos e 30 segundos de registro.

A Figura~\ref{fig:hr_devices} apresenta os registros de frequência cardíaca de ambos os dispositivos ao longo do protocolo. Observa-se que as transições entre as fases do protocolo estão bem refletidas em ambas as séries, evidenciando que a sincronização temporal foi eficaz. Além disso, as tendências gerais dos sinais são bastante consistentes, com o \textit{smartwatch} acompanhando de forma adequada a referência estabelecida pela cinta Polar H10, especialmente durante as fases de corrida e caminhada. No entanto, são notáveis algumas interrupções e variações abruptas nos dados do \textit{smartwatch}. Como mencionado anteriormente, por decisão do fabricante, o valor da frequência cardíaca (FC) é definido como zero quando a qualidade do sinal é considerada ruim — situação que ocorre, principalmente, em condições de maior movimento.

De forma geral, a ferramenta demonstrou capacidade de coletar e sincronizar os dados com boa qualidade, permitindo sua utilização em análises posteriores.

\section{Conclusão}

O trabalho apresentou a Colepp, uma ferramenta multiplataforma e de código aberto para coleta e sincronização de dados fisiológicos e de movimento provenientes de dispositivos vestíveis. Sua arquitetura permite integrar múltiplos sensores, alinhar temporalmente os sinais e exportá-los em formato padronizado, facilitando a criação de conjuntos de dados personalizados e de alta qualidade. Os testes realizados indicam que a solução é eficaz para aplicações como monitoramento cardíaco e reconhecimento de atividades humanas, contribuindo para pesquisas e o desenvolvimento de algoritmos mais robustos e generalizáveis.

\bibliographystyle{abbrv}
\bibliography{sample-base}

\end{document}